\documentclass[a4paper,fleqn,usenatbib]{mnras}
\usepackage{newtxtext,newtxmath}
\usepackage{cancel}

\usepackage[T1]{fontenc}
\usepackage{ae,aecompl}
\usepackage{multirow}
\usepackage{gensymb}
\usepackage{xcolor}
\usepackage{graphicx}	
\usepackage{amsmath}	
\newcommand{\teal}{\textcolor{black}}
\newcommand{\tred}{\textcolor{black}}

\newcommand{\tr}{\textcolor{black}}
\newcommand{\tl}{\textcolor{black}}

\title[Sch\"{o}nberg-Chandrasekhar limit]{\tl{The Sch\"{o}nberg-Chandrasekhar limit in presence of small anisotropy and modified gravity}}

\author[S. Chowdhury $\&$ T. Sarkar]{
Shaswata Chowdhury, \thanks{shaswata@iitk.ac.in}		Tapobrata Sarkar, \thanks{tapo@iitk.ac.in}
\\
Department of Physics, Indian Institute of Technology Kanpur, Kanpur 208016, India
}

\date{Accepted XXX. Received YYY; in original form ZZZ}

\pubyear{2015}

\begin{document}
\label{firstpage}
\pagerange{\pageref{firstpage}--\pageref{lastpage}}
\maketitle
	

\begin{abstract}
The Sch\"{o}nberg-Chandrasekhar limit in post main sequence evolution for stars of masses in the range 
$1.4\lesssim M/M_{\odot}\lesssim 6$ gives the maximum pressure that the stellar core can withstand, once the
central hydrogen is exhausted. It is usually expressed as a quadratic function of $1/\alpha$, with $\alpha$ being the
ratio of the mean molecular weight of the core to that of the envelope. 
Here, we revisit this limit in scenarios where the pressure balance equation in the stellar
interior may be modified, and in the presence of small stellar pressure anisotropy, that might arise due
to several physical phenomena. Using numerical analysis, we derive a three parameter dependent master formula for the limit, 
and discuss various physical consequences. As a byproduct, in a limiting case of our formula, we find that in the standard
Newtonian framework, the Sch\"{o}nberg-Chandrasekhar limit is best fitted by a polynomial that is linear, rather
than quadratic, to lowest order in $1/\alpha$.
\end{abstract}



\section{Introduction}

A main sequence star, after exhaustion of hydrogen in the central region, develops an isothermal helium core, 
surrounded by a radiative envelope, predominantly of hydrogen. Now, hydrogen in the core-envelope junction continues 
burning and adding mass to the core. However, the core mass fraction cannot exceed a certain upper limit $q_{max}$, known as the
Sch\"{o}nberg-Chandrasekhar (SC) limit as shown by \cite{HenChan}, \cite{SC}, following an earlier work by \cite{Gamow}. 
The SC limit is given by the textbook result $q_{max}\sim 0.37(1/\alpha)^2$ (\cite{CoxGiuli}, \cite{Kippenhahn}, \cite{CarrollOstlie}), 
where $\alpha = \mu_{c}/\mu_{e}$ corresponds to the ratio of mean molecular weight of the core ($\mu_c$) to that 
of the envelope ($\mu_e$), and is crucial in the analysis of post main sequence 
stellar evolution. It physically corresponds to the maximum admissible core pressure, which can withstand the pressure of 
the overlying layers in the envelope. 
When the core mass fraction reaches the SC limit, the hydrogen-depleted 
core begins to contract rapidly on a Kelvin-Helmholtz timescale\tl{, thus accelerating the stellar evolution}. The 
gravitational potential energy released in the process, leads to 
expansion of the stellar envelope, thus decreasing the effective 
temperature. This phase of very rapid redward evolution, known as the 
subgiant branch (SGB), on the Hertzsprung-Russel (H-R) diagram results 
in the Hertzsprung gap. This analysis is valid for stars with masses
$\tred{1.4}\lesssim M/M_{\odot}\lesssim 6$, as it is known that stars with masses less than $1.4M_{\odot}$ remain isothermal
above the SC limit (\cite{HHS}) and those with masses greater than $6M_{\odot}$ have core mass fractions 
that are \tred{higher} than this limit when central hydrogen is exhausted (see the discussion in \cite{ZZ}).

One of the basic ingredients used in deriving the SC limit is the pressure
balance equation inside a stellar object, which, in the Newtonian limit and assuming isotropy, is given by
$dP(r)/dr = - GM(r)\rho(r)/r^2$ with $P(r)$ and $\rho(r)$ being the pressure and density as a function of the radial 
coordinate $r$, $M(r)$ is the stellar mass up to $r$, and $G$ is Newton's gravitational constant. 
Suppose that this pressure balance equation is altered: then it is but natural that the SC limit is modified 
and that such modifications depend on the parameters characterising the alterations. In fact, these alterations are 
interesting and might be artefacts of important physical processes. Their importance lies in the fact that they
provide modifications of the SC limit within the realms of classical dynamics. 
For example, a star with mass $M/M_{\odot}\sim 1.4$ might have a partially degenerate core, and the
degeneracy pressure arising out of quantum effects, can play a crucial role in providing a higher value of 
the envelope pressure that can be supported by the core. As we will see later, a modification of the pressure 
balance equation can mimic this effect, to a small degree at the classical level, i.e., with an isothermal
non-degenerate core. Similarly for
stars with $M/M_{\odot}\sim 6$, such modifications might imply that a star that would have undergone
contraction according to the standard framework may not do so in the modified scenario, and vice versa. 
Since the SC limit is an estimate and not a precise measurement, it is difficult to quantify this 
statement further, but nonetheless provides an interesting astrophysical effect arising due to the
alteration of the stellar pressure balance equation. We recall that an analysis of the time spent by a star 
in the shell hydrogen burning stage due to the change in the SC limit was considered by \cite{Maeder}.

In this paper, we consider two such possible scenarios, namely modified
gravity and local pressure anisotropy, and the main contribution of this paper is to derive an analytic  
formula that provides the explicit dependence of the SC limit due to changes in the pressure balance equation 
in stellar interiors arising out of these two effects. 
First, we elaborate upon the main motivations for this study. Indeed, theories of gravity that involve extensions 
of the Einstein-Hilbert action have become immensely popular in the last few decades, as the search for the underlying 
mechanism of observed acceleration of cosmic expansion point towards unavoidable extensions to general relativity (GR) 
(see, e.g. the reviews in \cite{CliftonRev}, \cite{LangloisRev}, \cite{IshakRev}, \cite{KaseRev}). Here, we will be 
interested in the most general class of the so called ``beyond-Horndeski" models. To wit, 
out of the many possible modifications of GR, scalar-tensor theories (STTs), which incorporates scalar fields in 
the conventional Einstein-Hilbert action, are one of the best studied and most important ones (for an elaborate
treatment, see the monograph by \cite{fujimaeda}). Horndeski theories (\cite{Horndeski}), first discussed nearly five decades back, 
constitute the most general STTs, which are physical (i.e. the equations of motion 
are second-order and thus ghost-free). More recently these Horndeski theories have been generalized into the beyond-Horndeski 
class of theories (\cite{BH1}, \cite{BH2}) and these are free from unphysical ghost degrees of freedom 
in spite of exhibiting higher-order equations of motion. 

Now, any such modifications to GR, which are relevant at cosmological scales, need to be compatible with precise astrophysical 
tests as well. One therefore requires screening mechanisms, which screen the modified gravitational effects at 
astrophysical scales, thus recovering GR, while retaining the modifications at cosmological scales. The Vainshtein 
mechanism (\cite{Vainshtein}) is one of the most efficient screening mechanisms known to date 
(see, \cite{BabichevRev} for a review, see also \cite{JainKhoury}), \tred{which} recovers GR in the near regime through a non-linear effect. 
Importantly, a partial breaking of the Vainshtein mechanism (i.e. the break down of screening inside stellar objects) 
has been demonstrated in beyond-Horndeski theories, as first shown by \cite{KWY}. This leads to a modification 
of the pressure balance equation inside astrophysical objects, through an additive term depending on a dimensionless
parameter $\Upsilon$, which renormalises 
the Newton's constant that changes the strength of gravity inside a stellar object (\cite{SaksteinPRD}) and 
represents the effect of modified gravity. 

This fact provides an ideal laboratory for testing modified gravity theories and constraining 
them by astrophysical observations. This has received considerable attention of late in various settings, for 
example white dwarfs (\cite{Sakstein}, \cite{Jain}, \cite{Tapo1}), main sequence stars (\cite{SaksteinPRD}, 
\cite{Saito}, \cite{Tapo2}, \cite{Babichev}, \cite{SaksteinStrong}), cataclysmic variable binaries 
(\cite{Tapo3}, \cite{Tapo4}), \teal{very low mass objects (\cite{Wojnar2}, \cite{Wojnar3})}, \tr{giant planets (\cite{Wojnar4})}, etc. For recent reviews see \cite{Olmo1}, \cite{SaksteinRev}. 
One of the purposes of this paper is to establish the nature of the SC limit in a gravity 
theory belonging to the beyond-Horndeski class.

Interestingly, one can in fact envisage other important effects, which may at least partially modify the results of such 
modifications of gravity inside stellar objects. For example, stellar rotation effectively weakens gravity in the interior
of the star, and might in principle have a competing effect with those due to modified gravity. Another important 
issue is that of magnetic fields inside stellar objects. Indeed, since we are considering
stars with masses \tred{$1.4\lesssim M/M_{\odot}\lesssim 6$}, these effects, which are conveniently measured by the
quantity $P_{\perp} - P_r$ where $P_r$ is the radial pressure, and $P_{\perp}$ denotes the pressure in the 
orthogonal directions, will presumably be small (it vanishes in the isotropic case $P_{\perp} = P_r$). However, we should 
remember that the effects of beyond-Horndeski theories inside such a star is itself small, and hence it is important
to quantify these in any astrophysical considerations of such modified gravity theories. Now, it is known that rotation
or tidal effects or the presence of a magnetic field might in general cause pressure anisotropy inside a star.
Rotation, for example, causes a deformation of the stellar surface due to this.

What we will be interested in this paper is the presence of a 
stellar magnetic field that might \tred{induce} small anisotropy in the pressure. The topic has received considerable interest in the literature
in the context of stars with high magnetic fields, e.g. neutron stars or white dwarfs where the magnetic field can be
of the order of $10^{12}G$ or $10^7G$, respectively. Literature on magnetic fields in the intermediate
mass stars that we study here is relatively scarce (for a recent analysis of the time evolution of stellar 
magnetic fields in upper main sequence stars, see \cite{Landstreet}), and the topic has only recently started receiving 
attention (\cite{Quentin}, \cite{Takahashi}) as the importance of stellar magnetism for intermediate mass stars is becoming clearer. 
As discussed by \cite{Ferrer}, the presence of a magnetic field breaks rotational symmetry and hence induces an anisotropy
in the pressure. This anisotropy can be analytically determined in the case of strong magnetic fields, as was done by \cite{Ferrer}
and in a series of papers by \cite{Canuto1}, \cite{Canuto2}, \cite{Canuto3}. In the post main sequence stars that we consider,
the magnetic fields are weaker, and can maximally be $\sim 10^4 G$ at the stellar surface (see, e.g. \cite{Quentin}).  

Now, determining the exact nature of anisotropy for such small magnetic fields that we consider here 
is a daunting task and we are not aware of such an attempt till now, possibly because the effect is anyway small. 
However, as we have just mentioned, it is important in our context, as it provides a competing effect with that of
modified gravity. In what follows, we will proceed with the assumption 
that spherical symmetry is retained to a very good approximation, since the anisotropy due to the magnetic field is small. 
With this assumption, we can use a phenomenological model for anisotropy due to stellar magnetic fields. There are
two popular models in this context. \cite{HH} use the model $P_{\perp} - P_r \propto P_r$ while \cite{HS} use
$P_{\perp} - P_r \propto r^n$. As was argued by \cite{Tapo1}, the latter model is perhaps more suited towards
modelling rotational effects, and we will instead consider here $P_{\perp} - P_r = \beta(r) P_r$ as our model, with 
$\beta(r)$ being a dimensionless anisotropy parameter which in our model depends on the stellar radius, 
and will be chosen so that the anisotropy is small, a precise quantification of which will be given in section \ref{aniso}. 

\tl{The sharp SC limit described in this paper, which is indicative of the beginning of accelerated stellar evolution, is a consequence of the theoretical idealization of strict isothermality of the helium-rich core. However, using the stellar evolutionary code of \citet{paczynski69,paczynski70}, calibrated by \cite{Zdziarski}, it was shown by \cite{ZZ} that the accelerated stellar evolution phase is more appropriately characterised  by a range of fractional core masses, denoting a SC transition, rather than a single value of the core mass fraction denoting a strict SC limit.
Since the SC transition is inclusive of the SC limit, any changes in the SC limit due to physical phenomena should in principle reflect upon the SC transition as well. We will 
therefore consider a phenomenological model with its standard theoretical approximations, to obtain the effects of modified gravity and anisotropy on the SC limit. We emphasize from the outset that we are not constructing evolutionary tracks of stars, or creating a detailed representation of the stellar interior in modified gravity and anisotropic situations. These are important issues which deserve further study.}

What we then do in \tred{the main body of} this paper is to understand the SC limit in modified gravity, 
in the presence of small anisotropy. \tred{We derive the relevant equations and formulae in their non-dimensional 
forms in the isotropic case and in the presence of anisotropy, in Appendix \ref{isotropic} and \ref{anisotropic} respectively.} 
In particular, using numerical analysis, we will, at the end derive a master formula that expresses the SC limit as 
a function of $\alpha$, $\Upsilon$ and \tred{$\tau$, where $\tau$ is a non-dimensional constant quantifying 
the anisotropy parameter $\beta(r)$}. In addition, the reduced $\chi^2$ analysis used to achieve this shows that, 
at the lowest order, $q_{max}$ is best fitted when a linear function of $1/\alpha$ is added to 
the quadratic function of $1/\alpha$ in the Newtonian limit of the isotropic case,
rather than the sole quadratic function commonly used in the literature. 
Throughout this paper, we will present results on a core-envelope stellar model with 
an isothermal \tred{core} and an $n=3$ polytropic envelope. We have also considered a second model with an isothermal 
non-degenerate core, surrounded by a radiative envelope governed by Kramer's opacity law and a hydrogen 
burning shell at the core-envelope junction. We relegate discussion on this second model to the final
section \ref{disc}, as they give almost identical results.

Throughout the rest of this paper, the isotropic Newtonian limit of GR will be called the ``standard'' case, and 
deviations from this standard case should be obvious from the context.

\section{Stellar structure equations in beyond-Horndeski theories}
\label{sec1}

We begin by reviewing some standard facts about the modified gravity theory that we focus on here. 
Recall that in a generic situation, the stress-energy tensor inside a spherically symmetric stellar object
possessing pressure anisotropy is given by $T^{\mu}_{\nu} = {\rm diag}(-\rho c^2, P_{r}, P_{\perp}, P_{\perp})$, 
where $c$ is the speed of light. To study these in beyond-Horndeski theories, following \cite{KWY}, 
one considers a perturbation about a static Friedman-Robertson-Walker universe corresponding to a spherical overdensity
in the Newtonian limit, given by the metric 
\begin{align}
ds^2 = &-\left(1 + 2\Phi(r)\right)c^2dt^2 \nonumber \\
&+ \left(1-2\Psi(r)\right)\left[dr^2 + r^2\left(d\theta^2 
+ \sin^2\theta d\phi^2\right)\right]~.
\end{align}
Here, $\Phi(r)$ and $\Psi(r)$ are metric potentials ($\Phi(r),\Psi(r) \ll 1$), with the former being the 
Newtonian gravitational potential. Covariant conservation of stress-energy tensor, 
i.e., $D_{\mu}T^{\mu\nu}=0$ (with $D_{\mu}$ being the covariant derivative), then gives 
\begin{equation}
\frac{dP_{r}}{dr} = -\rho c^2\frac{d\Phi}{dr} +\frac{2}{r}\left(P_{\perp}
-P_{r}\right)\left(1-r\frac{d\Psi}{dr}\right)=0
\label{TOVA2}
\end{equation}
The equations governing the metric potentials, in these theories were derived by \cite{KWY} and \cite{SaksteinPRD},
and are given by 
\begin{equation}
\frac{d\Phi}{dr} = \frac{GM(r)}{c^2r^2} + \frac{\Upsilon}{4}\frac{G}{c^2}\frac{d^2M(r)}{dr^2}~~,~~
\frac{d\Psi}{dr} = \frac{GM(r)}{c^2r^2} - \frac{5\Upsilon}{4}\frac{G}{c^2 r}\frac{d M(r)}{d r}~,
\label{phipsider}
\end{equation}
where $\Upsilon$ is a dimensionless parameter introduced in beyond-Horndeski theories and arises in an 
effective field theory of dark energy. This is the only free parameter in these theories that carries
the information of modifications to standard gravity theories. Also,  
$dM(r)/dr=4\pi r^{2}\rho(r)$. Now, the modified pressure balance
equation can be obtained by substituting Equation (\ref{phipsider}) in Equation (\ref{TOVA2}) and
taking the Newtonian limit (\cite{Tapo1}). The isotropic case with $P_{r} = P_{\perp}$ will be discussed
in the next section \ref{modSC}, while the general anisotropic case will be the subject of
section \ref{aniso}. 

For completeness, we note that the radiative transfer equation and energy transport condition, 
not being dependent on the theory of gravity, remain unaltered, and are given respectively by
(\cite{SaksteinPRD})
\begin{equation}
\frac{dT(r)}{dr}=-\frac{3}{4a}\frac{\kappa}{T(r)^3}\frac{\rho(r) L(r)}{4\pi r^2}~,
\label{RTE}
\end{equation}
\begin{equation}
\frac{dL(r)}{dr}=4\pi r^{2}\tr{\rho(r)}\epsilon(r)~,
\label{ETC}
\end{equation}
with $L(r)$ and $T(r)$ \tred{being} the luminosity and temperature at a radial distance $r$. 
In the above, $\kappa$ is the opacity, $\epsilon(r)$ is the \tr{energy released per unit mass per unit time}, and $a$ is 
the radiation-density constant. These, along with the modified pressure balance equation forms the 
mathematical ingredients that we will require. 

Now, we would be specifically interested in models with an isothermal core surrounded by an envelope 
in radiative equilibrium. This effectively turns Equation (\ref{ETC}) into a mathematical identity, i.e.,
$dL(r)/dr=0$. Hence, from this point onwards, we drop Equation (\ref{ETC}) from our discussions and take $L(r)=L$, and
impose the condition $L=0$ inside the isothermal core and non-zero in the radiative envelope. 
Solutions to the equations discussed above must fulfil the following boundary conditions at $r=0$ (center)
and at $r=R$ (the stellar surface) : 
$$
M(0)=0~,~T(0)=T_{c}~,~P(0)=P_{c}~;~~ M(R)=M~,~T(R)=0=P(R)~
$$
where $M$ and $R$ are the stellar mass and radius respectively, while $T_{c}$ is the isothermal core temperature 
and $P_{c}$ the central pressure.

The continuity of stellar structure variables like mass, pressure \tred{and radius} at the core-envelope 
junction are ensured by fitting the core and envelope solutions through homology invariants $U$ and $V$ defined as
\begin{equation}
U=\frac{d \ln M(r)}{d \ln r}~~~~,~~~~V=-\frac{d \ln P(r)}{d \ln r}~
\label{UV}
\end{equation}
They satisfy the fitting condition
\begin{equation}
U_{fe}=\frac{1}{\alpha} U_{fc}~,~~V_{fe}=\frac{1}{\alpha} V_{fc}~.
\label{fit}
\end{equation}
\tred{Here, the subscript ``fc" denotes the fitting point, i.e., the core-envelope junction, approached from the 
core side, while the subscript ``fe" denotes the same when approached from the envelope side.}
In this paper, we will use the $U-V$ plane analysis. We note that the SC limit can be obtained approximately
by the virial theorem, as detailed in most textbooks (see also the chapter by Stein in \cite{Stein}). 

\section{Modification to the SC Limit in the isotropic case}
\label{modSC}

With these mathematical preliminaries, we are ready to start the discussion on effects of modified gravity
on the SC limit in the isotropic case. As a warm-up 
exercise, let us first understand what an analytic formalism reveals in this setting. 
Using the ideal gas equation and quasi-constant matter density 
(i.e., $d\rho(r)/dr=0$), from standard methods (see, e.g. \cite{CarrollOstlie}),
we arrive at the expression for maximum pressure $P_{\rm{core}}$ exerted by the isothermal core at the 
core-envelope junction as
\begin{equation}
P_{\rm{core}}\big|_{max}=\frac{c_1}{\overline{G}^{3}M_{c}^{2}}
\left(\frac{k_BT_{c}}{\mu_{c}m_{H}}\right)^{4}~,
\label{Pisomax0}
\end{equation}
where $M_{c}$ is the core mass,
$c_1$ is a numerical factor, \tred{$m_{H}$ is the mass of a hydrogen atom, and $k_B$ the Boltzmann constant.} 
Also, $\overline{G}=G(1+\frac{3\Upsilon}{2})$ is the renormalised Newton's gravitational constant and carries the information
about the modification of gravity. From Equation (\ref{Pisomax0}), we see that higher value of $\Upsilon$ lowers 
the maximum value of $P_{\rm{core}}$. This is expected, since it is known that increasing $\Upsilon$ generally
leads to a weakening of gravity inside a stellar object (\cite{SaksteinPRD}, \cite{Sakstein}).
From similar assumptions as above we obtain the following expression for the pressure $P_{e}$ at the 
core-envelope junction, exerted by the overlying layers of the envelope.
\begin{equation}
P_{e}=\frac{c_2}{\overline{G}^{3}M^{2}}\left(\frac{k_BT_{c}}{\mu_{e}m_{H}}\right)^{4}~,
\label{Penvfinal0}
\end{equation}
where $c_2$ is a numerical factor. 
We see that $P_{e}$ monotonically decreases with increase in $\Upsilon$ value, for a given $M$, $T_{c}$, and $\mu_{e}$. 
As before, this can \tred{be} attributed to the weakening of gravity inside a stellar object with increasing $\Upsilon$.
From the stability criterion $P_{e}\leq P_{\rm{core}}|_{max}$, and restoring numerical factors, 
we obtain the SC limit as $q_{max}=0.25\left(\frac{1}{\alpha}\right)^{2}$, which is close to the formula
quoted in the introduction. Naively, 
the limit does not depend upon the modified gravity parameter $\Upsilon$ as the factors of ${\overline{G}}$
cancel out. The most important takeaway message from this analytic study, is that, 
the  SC limit seems to be unaffected by modified gravity. At this point, one is 
free to argue that such an analytic computation has a large number of approximations and may not have
captured the effect of modified gravity on the SC limit. In fact, this is exactly what we will be \tred{demonstrating} below. 
Once we relax many of the analytical approximations, and treat the problem semi-analytically, 
significant changes appear in the SC limit, due to modified gravity.

Now, in our semi-analytic study, we derive the effects of modified gravity on SC limit in the isotropic 
case. There are two models popular
in the literature for such a study: A) a model with isothermal core and $n=3$ polytropic envelope, and 
B) a model with isothermal core and radiative envelope governed by Kramer's opacity law. The \tred{generic} numerical 
methodology essentially involves the following steps (\cite{Schwarzschild}) : 

\begin{itemize}
\item Integrating the pressure balance, the mass conservation and the radiative equilibrium equations from the centre outwards, 
with the initial conditions at the core, we obtain a family of core solutions parameterised by $T_{c}$ and $P_{c}$.
\item Integrating these equations from the surface inwards, with the corresponding initial conditions at the stellar surface, 
we obtain a family of envelope solutions parameterised by $M$, $R$ and $L$.
\item Fitting of the two family of solutions, through \tred{homology} invariants $U$ and $V$. 
The fitting point corresponds to \tred{the} core-envelope junction. $U$ and $V$ are defined in Equation (\ref{UV}), which needs to 
satisfy the relation given in Equation (\ref{fit}) at the fitting point, which ensure continuity of pressure, mass and 
radius across the core-envelope junction.
\end{itemize}

\subsection{SC limit with an isothermal core and an $n=3$ polytropic envelope}
\label{ModelA}

\begin{figure*}
\begin{tabular}{cc}
\includegraphics[width=80mm]{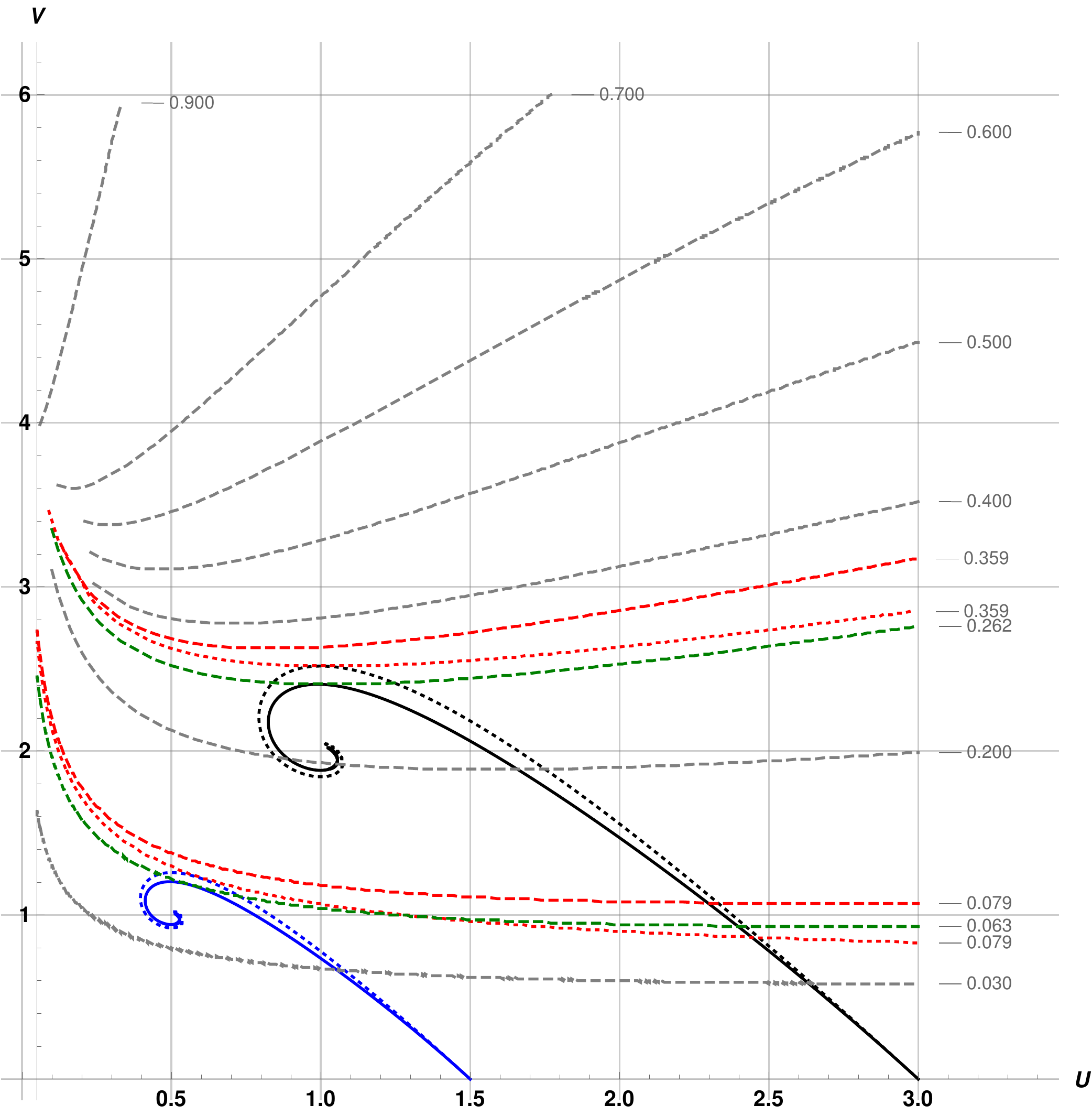} & \includegraphics[width=80mm]{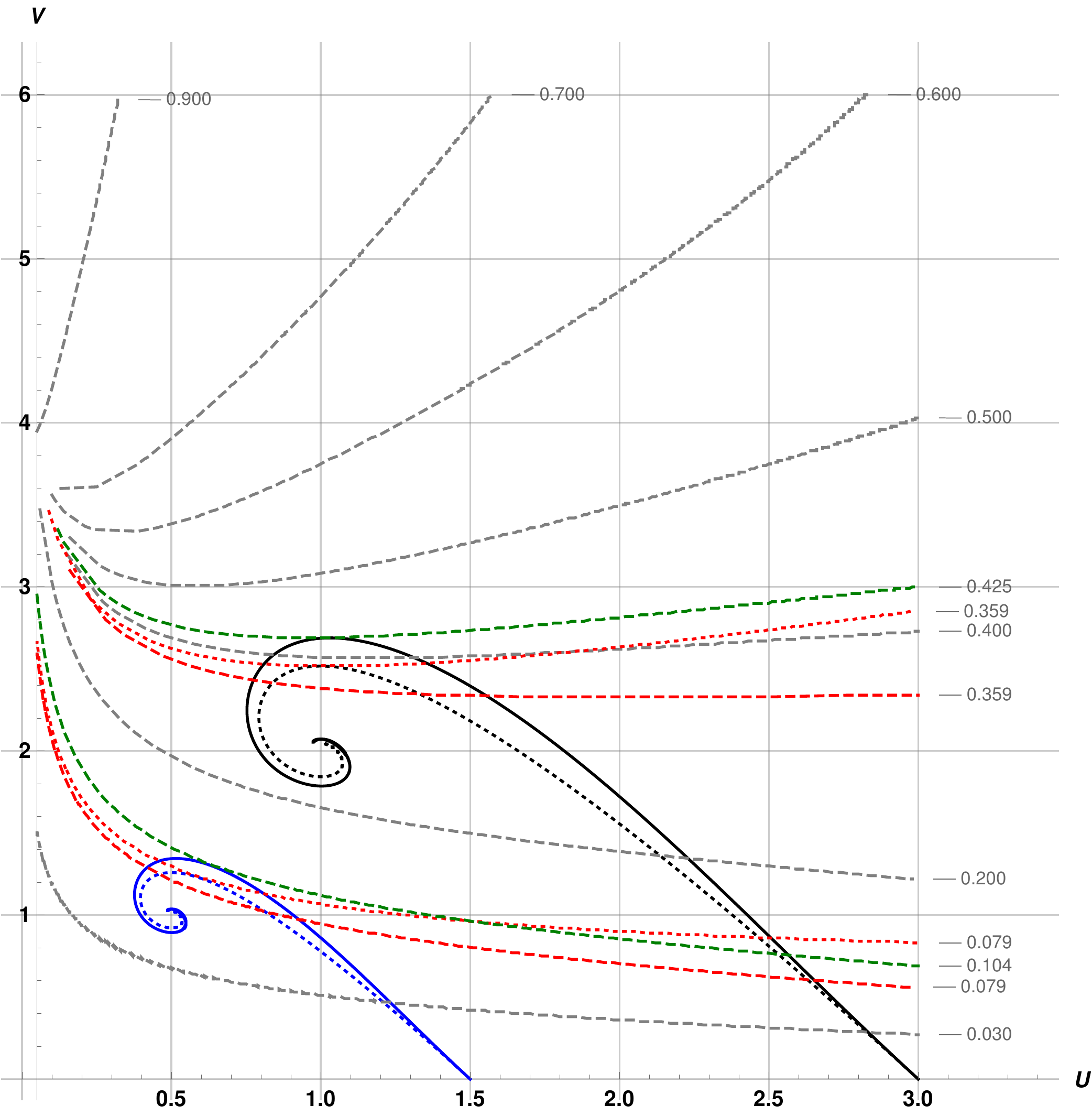} \\
(a) $\Upsilon=0.25$ & (b) $\Upsilon=0.25$  \\[4pt]
\end{tabular}
\caption{\tred{$U-V$ plane for $\Upsilon=0.25$ (a) and $\Upsilon =-0.25$ (b). 
The \teal{black solid} spiral represents isothermal core with $\alpha=1$, while the dotted one represents 
the same for $\Upsilon=0$. The \teal{blue solid} spiral represents isothermal core with $\alpha=2$, while the dotted 
one represents the same for $\Upsilon=0$. The dashed curves represent contours of core mass fraction for $n=3$ 
polytropic envelope; the green curves denoting the corresponding SC limits. The \teal{red dotted} curves denote the 
contours of core mass fraction corresponding to the SC limits for $\Upsilon=0$, while the \teal{red dashed} ones denoting 
the same core mass fraction contours but for the given value of $\Upsilon=0.25$.}}
\label{figure1}
\end{figure*}

Using the Eddington standard model, a radiative envelope can be well approximated by a polytrope of finite 
polytropic index ($n=3$) (see, e.g. \cite{CarrollOstlie}). This is also what \cite{HenChan} had chosen in their work. 
We would be specifically following \cite{BallTout} and work in the space of homology invariants, i.e., the $U-V$ plane
(for analytical models of the SC limit using polytropic approximations, see \cite{Beech}, \cite{Eggleton}).
Firstly, for a given $\Upsilon$, we solve the non-dimensionalised stellar structure equations in the isothermal core.
With $p^{*}$, $q^{*}$ and $x^{*}$ corresponding to the non-dimensional pressure, mass and radial coordinate 
inside the core \tred{(see Appendix \ref{isotropic})}, we have
\begin{align}
&\frac{dp^{*}(x^{*})}{dx^{*}}=-\frac{q^{*}(x^{*})p^{*}(x^{*})}{{x^{*}}^{2}}-\frac{\Upsilon}{4}p^{*}(x^{*})
\frac{d^{2}q^{*}(x^{*})}{d{x^{*}}^2}~,\nonumber \\
&\frac{dq^{*}(x^{*})}{dx^{*}}=p^{*}(x^{*}){x^{*}}^{2}~,
\label{MCCA}
\end{align}
\tred{as the non-dimensionalised pressure balance and mass conservation equations, respectively.}
At $x^{*}=0$ (i.e., the stellar center), we have $q^{*}(0)=0, p^{*}(0)=1$ \tred{as the boundary conditions. 
Equation (\ref{MCCA}) along with these} boundary conditions yield
the homology invariants $(U,V)$ at every point of the isothermal core for a given value of $\Upsilon$. 
In the process, a spiral curve in the $U-V$ plane is obtained, \tred{which we call the core curve}. We now contract this curve, 
by the factor of $\alpha$, which corresponds to the density jump at the core-envelope junction. This shifting is essential for the 
subsequent fitting of a polytropic envelope to the isothermal core.
Next, we consider a point $(U_{0},V_{0})$, in the same $U-V$ plane, to correspond to the core-envelope junction. 
Using the above $\Upsilon$ value, we integrate the non-dimensionalised pressure balance and mass conservation 
equations inside the polytropic envelope \tred{from this particular point up to the stellar surface, subject to the appropriate 
boundary conditions at the core-envelope junction}. With $\theta$, $\phi$ and $\xi$ being the 
non-dimensional variables \tred{in the envelope,} associated to density, mass and radial coordinate respectively  
\tred{(see Appendix \ref{isotropic})}, 
the non-dimensionalised pressure balance and mass conservation equations are given respectively as,
\begin{align}
&\frac{d\theta(\xi)}{d\xi}=-\frac{\phi(\xi)}{\xi^2}-\frac{\Upsilon}{4}
\left(2\theta(\xi)^{n}\xi+\theta(\xi)^{n-1}\xi^{2}n\frac{d\theta(\xi)}{d\xi}\right)~, \nonumber \\
&\frac{d\phi(\xi)}{d\xi}=\xi^{2}\theta(\xi)^n~.
\label{MCCpoly}
\end{align}
\tred{At the core-envelope junction $\xi=\xi_0,~~\theta(\xi_{0})=1,~\phi(\xi_{0})=\phi_{0}$ serves as the boundary condition.} 
Here, we have defined 
\begin{equation}
\xi_{0}=\left(\frac{V_{0}}{\frac{(n+1)}{U_{0}}+\Upsilon\left(\frac{(n+1)}{2}-\frac{nV_{0}}{4}\right)}\right)^{1/2}~,
~~\phi_{0}=\frac{\xi_{0}^{3}}{U_0}
\label{startpointUV}
\end{equation}
\tred{The stellar surface corresponds to the first zero of $\theta(\xi)$.}
Let the first zero of $\theta(\xi)$ correspond to $\xi=\xi_{1}$. According to the definition of 
$\phi$ in Equation (\ref{Trans3}), the core mass fraction $q_{c}$ is the ratio $\phi(\xi_{0})/\phi(\xi_{1})$. 
For each and every point in the $U-V$ plane, we calculate the corresponding $q_{c}$ values, thus giving us the
function $q_{c} (U,V)$. We plot the contour lines for this function, in the $U-V$ plane for 
\tred{the given value} of $\Upsilon$. \tred{We perform this for different values of $\Upsilon$, in 
Figures \ref{figure1}(a) and \ref{figure1}(b).} 

From these figures, it is seen that, \tred{for any given value of $\alpha$}, the isothermal core admits a maximum $V$, 
which corresponds to the maximum pressure, that can be supported by the same, for a given mass (see \cite{BallTout}). 
The existence of such maxima, ensures that the contours of core mass fraction $q_{c}$, above a certain maximum value 
$q_{max}$, do not admit any intersection with the isothermal core solution \tred{for a given $\alpha$ value}. 

For example we find that for the standard case, i.e., with $\Upsilon=0$, the 
contour lines corresponding to $q_{c}>0.359$ \tred{do} not admit any intersection with the isothermal 
core solution for $\alpha=1$. Hence here $q_{max}=0.359$ for $\alpha=1$ and similarly we find that 
$q_{max}=0.079$ for $\alpha=2$. This value of $q_{max}$ then corresponds to the SC limit for the given value
of $\alpha$. Also, the limit depends on the particular value of the polytropic index. 
The case $\Upsilon\neq 0$ is similarly dealt with, as shown in Figures \ref{figure1}(a) and \ref{figure1}(b).
The salient features of this analysis is summarised below

For $\Upsilon>0$, we have weakening of gravity inside the stellar object, thus leading to a decrease in 
the maximum pressure, that can be supported by the isothermal core. This is effectively captured by the 
lowered value of maximum $V$, for the isothermal core, as compared to the case for $\Upsilon=0$, see Figure \ref{figure1}(a). 
The contour lines also show different behaviour compared to the $\Upsilon=0$ case. These have effectively moved upwards, 
so that a given point $(U,V)$, in the $U-V$ plane will now correspond to a lower value of the core mass fraction, 
as compared to the value for $\Upsilon=0$\tred{; in Figure \ref{figure1}(a) compare the red dotted and dashed curves, 
corresponding to the contours of same core mass fraction but the former for $\Upsilon=0$, while the latter is
for $\Upsilon=0.25$}. \tred{For} $\Upsilon = 0.25$, we find that $q_{max}=0.306$ for $\alpha=1$ and $q_{max}=0.063$ for $\alpha=2$.
For this value of $\Upsilon$, there is thus a $15\%$ decrease in the SC limit for $\alpha=1$, while a $20\%$ 
decrease in the same for $\alpha=2$ when compared with the standard case.

For $\Upsilon<0$, we have strengthening of gravity inside the stellar object, thus leading to an increase 
in the maximum pressure, that can be supported by the isothermal core. This is effectively captured by the increased 
value of maximum $V$, for the isothermal core, as compared to the standard case, see Figure \ref{figure1}(b). 
Here, not surprisingly, the contour lines have been shifted downwards\tred{; in Figure \ref{figure1}(b) 
compare the red dotted and dashed curves, corresponding to the contours of same core mass fraction with 
the former for $\Upsilon=0$, while the latter is for $\Upsilon=-0.25$}. \tred{For a typical value of} $\Upsilon=-0.25$, we find that
$q_{max}=0.425$ for $\alpha=1$ and $q_{max}=0.104$ for $\alpha=2$, leading to an $18\%$ increase in the SC limit 
for $\alpha=1$, and a $32\%$ increase for $\alpha=2$ compared to the $\Upsilon=0$ case.
\teal{In this context, we recall that modifications of the SC limit for rotating stars were considered by
\cite{Maeder}, who found significantly less changes for uniform rotation.}

As an upshot of the above discussion, we obtain a \tred{quartic fitting formula for the SC limit, 
$q_{max}=\sum_{x,y}C_{x y}(1/\alpha)^{x}\Upsilon^{y}$, with $x+y\leq4$. $C_{xy}$ are numerical coefficients as listed in 
the following Table \ref{modgravformula}.} 
\begin{table}
\caption{List of coefficients $C_{xy}$}
\begin{tabular}{|c|c|c|c|c|c|}
\hline
$x \backslash y$ & 0 & 1 & 2 & 3 & 4\\
\hline
$0$ & $-$ & $-0.018$ & $-0.410$ & $-0.174$ & $0.100$\\
$1$ & $0.136$ & $0.589$ & $1.50$ & $0.084$ & $-$\\
$2$ & $-0.544$ & $-2.00$ & $-1.02$ & $-$ & $-$\\
$3$ & $1.56$ & $1.20$ & $-$ & $-$ & $-$\\
$4$ & $-0.793$ & $-$ & $-$ & $-$ & $-$\\
\hline
\end{tabular}
\label{modgravformula}
\end{table}	
We have also evaluated the $\Upsilon$-dependent correction to the pressure at the core envelope junction. 
Writing this pressure as $P_{ce} = P_0 + \Delta P_{ce}(\Upsilon)$ where $P_0$ is the pressure with $\Upsilon=0$,
we find that \teal{$\Delta P_{ce}(\Upsilon)$} is best fitted by a quintic \teal{polynomial} in $\Upsilon$ and restoring dimensions, we find that
\begin{align}
\Delta P_{ce}(\Upsilon)/10^{16} ~({\rm dynes/cm^2})= -1.11\Upsilon &+ 1.11 \Upsilon^2 - 1.94 \Upsilon^3 \nonumber \\
&+ 
4.99 \Upsilon^4 - 4.27 \Upsilon^5~.
\label{fitdeltaP}
\end{align}
From Equation (\ref{fitdeltaP}), it is readily seen for example that values of $\Upsilon \sim$ a few times $10^{-1}$ can 
cause a significant change in $P_{ce}$, of the order of $P_0$ ($P_0$ can be estimated to be $\sim 5\times 
10^{15}~{\rm dynes/cm^2}$ from \cite{Blackler}). 
As mentioned in the introduction, such changes can alter the time that a star spends in the shell hydrogen burning stage.
Here we have provided the viability of such a process in modified gravity. \tl{A more quantitative analysis will be provided
at the end of the next section.}

\section{SC limit in modified gravity with small anisotropy}
\label{aniso}

As discussed in the introduction, we adopt a simple model for the pressure anisotropy given by 
$P_{\perp}-P_{r} = \beta(r) P_{r}$, so that the hydrostatic equilibrium condition following from Equation (\ref{TOVA2}) 
in the Newtonian limit becomes (\cite{Tapo1})
\begin{equation}
\frac{dP_{r}}{dr} = - \frac{GM(r)\rho(r)}{r^2} - \Upsilon\left(\frac{G\rho(r)}{4}\right)
\frac{d^2M(r)}{dr^2} + \frac{2}{r}\beta(r)P_{r}~,
\label{TOVA}
\end{equation}
where $\beta(r)$ is a dimensionless parameter measuring the strength of anisotropy. To proceed further, 
we will need to specify a form for $\beta(r)$. Various forms of this function of the radius is possible, and
we will adopt a model in which it is a polynomial in the radial distance. Now, in the case of polytropic stars, 
it was shown by \cite{Tapo1} that considerations near the centre of the star imposed restrictions on 
such a polynomial, so that one can work with $\beta(r)\sim r^2$. In our case, we will
adopt this as a phenomenological model. On the core side, we will hereafter choose  
$\beta(r)={\bar\tau}(r/R)^2$, where ${\bar\tau}$ is a dimensionless constant 
(see the discussion after Equation (\ref{HECAaniso})) and a similar choice is done for the envelope, as
detailed in Appendix \ref{anisotropic}. 

Now we will need to quantify what we mean by small anisotropy.
By this, we will demand that $P_{\perp}-P_{r}$ at a given radius will be maximally about an order of
magnitude less than the radial pressure at that point. This is clearly true near the stellar surface, where 
$P_r$ drops to zero, otherwise, we will require $\beta(r)$ to have a maximum value of $0.1$. 
Near the centre of the star, $\beta(r)$ is small, due to the smallness of $r/R$. Using the formulae described in 
Appendix \ref{anisotropic}, we find that as a function of radius, \tr{$P_{\perp}-P_{r}$} rises from the core and 
reaches its maximum value at the core-envelope junction whereafter it falls off again. At the junction, 
with $\alpha=1$ for example, we find that with $\tau=0.01$, $\beta=5.6\times 10^{-2}$, thus satisfying 
our criterion for smallness of anisotropy. 

The numerical analysis here
is more intricate \tred{than} what was outlined in the previous section \ref{ModelA}, and involves matching of the 
anisotropy parameter of the model at the core-envelope junction \tred{(see Appendix \ref{anisotropic})}. 
\begin{figure*}
	\begin{tabular}{cc}
		\includegraphics[width=83mm]{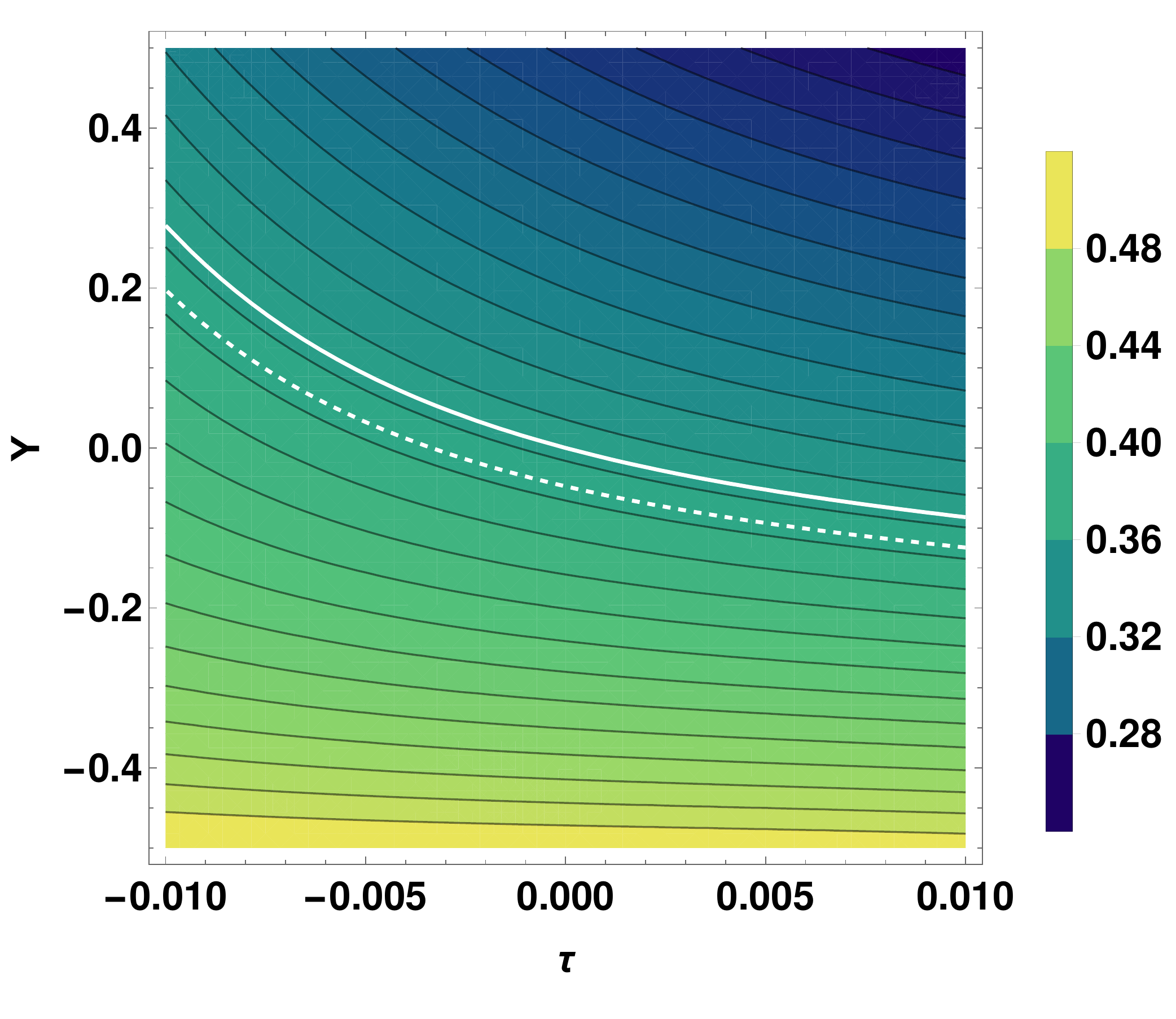} & \includegraphics[width=87mm]{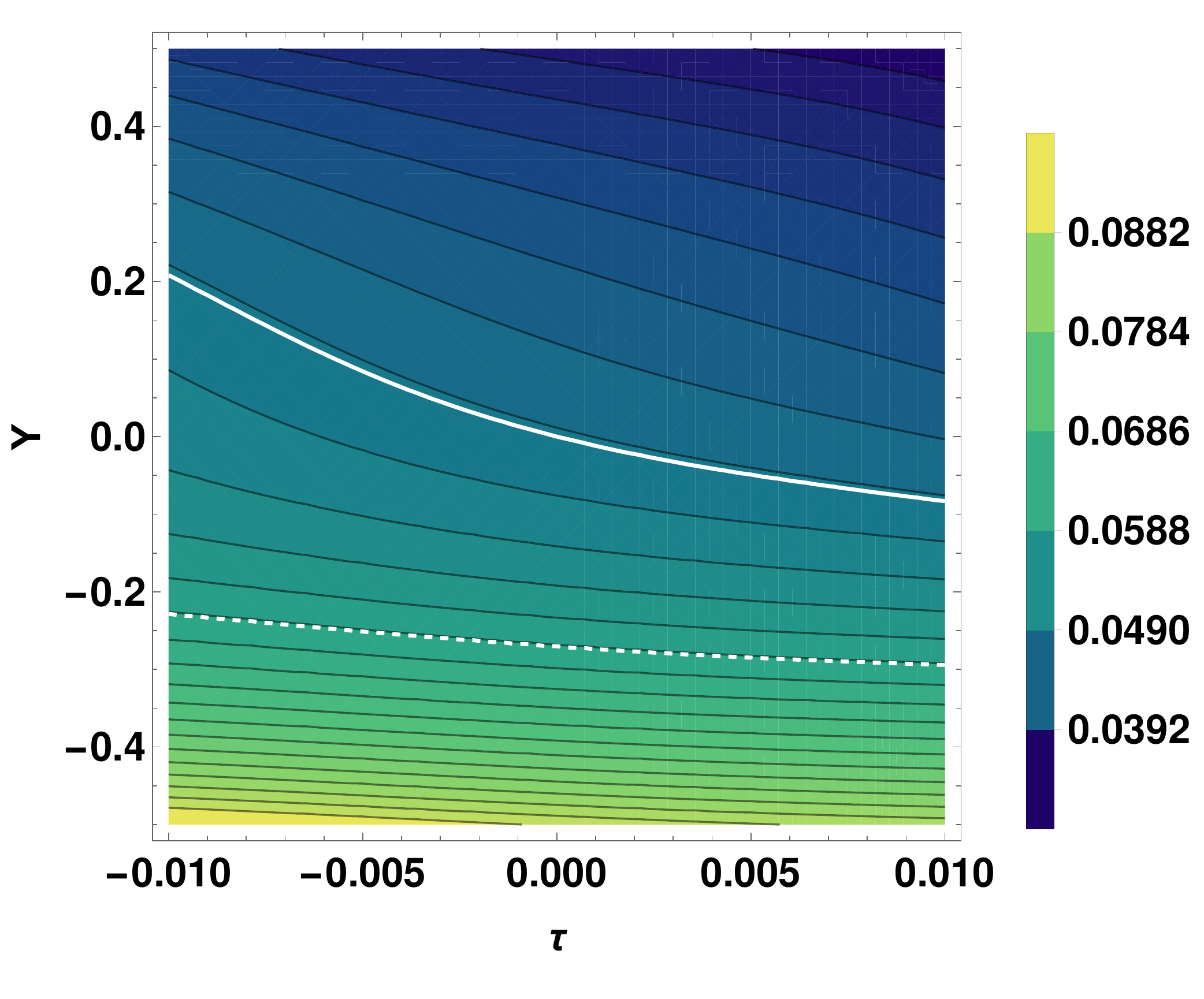} \\
		(a) $\alpha=1$ & (b) $\alpha=2.5$  \\[4pt]
	\end{tabular}
	\caption{Contours of the SC-limit for $\alpha=1$ (a) and $\alpha=2.5$ (b) from 
		the 4th order master formula. Any particular 
		contour specifies all admissible tuples of the modified gravity parameter $\Upsilon$ and the non-dimensional 
		anisotropy constant $\tau$ ($\tau$ quantifies the anisotropy parameter $\beta$, see Appendix \ref{anisotropic}), 
		for which one obtains identical SC limit. The \teal{white dashed} contour \tr{corresponds to} $q_{max}=0.37(1/\alpha)^2$.
		The \teal{white solid} contour \tr{corresponds to} the $\Upsilon=\tau=0$ limit of our master formula.}
	\label{figure2}
\end{figure*}

There are a few numerical subtleties related to the introduction of anisotropy. \tred{For negative anisotropy parameter values}, 
we observe that at a larger radial distance away from the center, the core pressure becomes very small, which makes the 
core curve rise upwards, irrespective of the $\Upsilon$ value chosen \tred{(See expression for $V$ in terms of the core variables in Appendix \ref{anisotropic})}. 
Since the junction pressure is usually taken to be a few orders of magnitude lower than the central pressure, 
we terminate the core solution at the point where pressure becomes $10^{-2}$ times the central pressure. This prevents 
the solution from blowing up. However, \tred{for positive anisotropy parameter values}, the pressure profile inside the core shows 
oscillations after the initial dip from the maximum central value and then saturates to a non-zero value at a 
larger radial distance, rather than going down to zero. The reason is the presence of competing terms in the pressure 
balance equation of the core \tred{(See Appendix \ref{anisotropic})}. This eventually leads the core curve to take negative $V$ values, which is unphysical 
considering the fact that pressure inside the star can never increase with the radial distance. We thus terminate the core 
solution at the point where the pressure starts rising. In our numerical analysis, both these truncation conditions are 
implemented simultaneously. We have checked that the results obtained from the truncation condition are in 
conformity with the existing results given in \cite{BallTout}, \tred{for the standard case}.

We obtain the SC limit for different values of $\alpha$, corresponding to several values of $\Upsilon$ and $\tau$, 
where $\tau$ is a non-dimensional constant quantifying the anisotropy parameter $\beta$ (see Appendix \ref{anisotropic}). 
All the three independent parameters are varied well within the typical admissible range; $\alpha$ varies from $1$ 
(homogeneous composition) to $2.5$ (completely ionised \tl{h}elium core with completely ionised \tl{h}ydrogen in the envelope), 
$\Upsilon$ varies from $-0.5$ to $0.5$, which is the typical range of $\Upsilon$ obtained from the astrophysical probes 
of modified gravity theories (see \cite{Tapo2}), $\tau$ varies from $-0.01$ to $0.01$ (see \cite{Tapo1}). Using a 
reduced $\chi^2$ analysis on our numerically obtained data, we obtain a \tred{quartic master formula of the SC limit} 
\tred{$q_{max}=\sum_{x,y,z}C_{x yz}(1/\alpha)^{x}\Upsilon^{y}\tau^{z} $}, with \tred{$x+y+z \leq 4$ and $C_{xyz}$ being 
numerical coefficients}. This is a lengthy formula, and it is best to 
show the results graphically. Our main findings here are summarised in Figures \ref{figure2}(a) and (b), 
where we show the contours of the SC limit for two \tred{different} values of $\alpha = 1$ and $2.5$, respectively. 
For a given $\alpha$, all the stars with particular values of ($\Upsilon$, $\tau$) tuples, constituting any 
given contour, will correspond to the same SC limit. The results from these plots are not to be extrapolated 
beyond the valid range of the parameters mentioned above. The \teal{white dashed} contour \tr{corresponds to} the \tr{conventional} SC limit $q_{max}=0.37(1/\alpha)^2$, \tr{mentioned in the introduction}. Note that the SC limit value for the standard case, as derived from our master 
formula, \tr{to which the \teal{white solid} contour corresponds,} differs from the \tr{conventional} SC limit $0.37(1/\alpha)^2$ by $\sim 2.8\%$ for $\alpha=1$ and 
by $\sim 20\%$ for $\alpha=2.5$. \tr{The reason for such differences lies in the fact that the \tr{conventional} formula contains only a quadratic term, while the standard limit of our quartic master formula contains linear, cubic, and quartic terms in addition to a quadratic one. Moreover, since the contribution from the linear term is enhanced for larger $\alpha$ values, such differences are observed to be larger for larger $\alpha$ values.}

Finally, we will present the dependence of the SC limit with the anisotropy parameter (with $\Upsilon = 0$), analogous
to what was obtained in the isotropic case in Table \ref{modgravformula} of subsection \ref{ModelA}. Writing
$q_{max}=\sum_{x,y}\bar{C}_{x y}(1/\alpha)^{x}\tau^{y}$, with $x+y \leq 4$ \teal{and $\bar{C}_{xy}$ 
denoting numerical coefficients}, our results here are tabulated in Table \ref{modgravformula4}.
\begin{table}
\caption{List of coefficients $\bar{C}_{xy}$}
\begin{tabular}{|c|c|c|c|c|c|}
\hline
$x \backslash y$ & 0 & 1 & 2 & 3 & 4\\
\hline
$0$ & $-$ & $-0.061$ & $104$ & $4.77\times10^3$ & $1.57\times10^5$\\
$1$ & $0.117$ & $-0.662$ & $-442$ & $\tl{-}7.61\times10^3$ & $-$\\
$2$ & $-0.450$ & $-0.569$ & $400$ & $-$ & $-$\\
$3$ & $1.42$ & $-1.39$ & $-$ & $-$ & $-$\\
$4$ & $-0.732$ & $-$ & $-$ & $-$ & $-$\\
\hline
\end{tabular}
\label{modgravformula4}
\end{table}        

\subsection{\tl{Physical consequence of a change in the SC limit}}
\label{physical}
\tl{One of the physical consequences of a change in the SC limit $q_{max}$ would be on the time $t_{shell}$ spent by an intermediate mass star in the shell hydrogen burning phase. 
An approximate order of magnitude relation between the relative change in SC limit and the altered shell burning lifetime $t_{shell}^{*}$ was 
given by \cite{Maeder} as
\begin{equation}
\frac{t_{shell}^{*}}{t_{shell}} \sim 1 + \frac{\Delta q_{max}}{q_{max}}\Big(\frac{\bar{L}}{L_{shell}}\Big)\Big(\frac{t_{MS}}{t_{shell}}\Big)~,
\label{lifetime}
\end{equation}
where $\Delta q_{max}$ represents a change in $q_{max}$ (due to any reason such as modified gravity, anisotropy, or stellar rotation) and $t_{MS}$ corresponds to the time spent during the main sequence phase. Also, $\bar{L}$ is the average luminosity during the main sequence and $L_{shell}$ is the stellar luminosity during the shell burning phase. Following \cite{Maeder}, we choose 
as an illustration $t_{shell}/t_{MS} \sim 0.042$ from \cite{Iben1965} as a representative value for a $3M_{\odot}$ intermediate mass star, and $L_{shell}/\bar{L} \sim 1.7$.}

	\tl{In order to simplify the discussion, we first focus on the isotropic case $\tau = 0$. Then, for a given value of $\alpha$, a positive 
$\Upsilon$ value leads to a decrease in the SC limit (see section \ref{ModelA}) so that $\Delta q_{max}$ in Equation (\ref{lifetime}) is negative. 
Physicality requires that the left hand side of Equation (\ref{lifetime}) has to be positive, and using the values given in the last paragraph, this 
translates to an upper bound $\Upsilon \lesssim 0.12$ for $\alpha=1$ and 
$\Upsilon \lesssim 0.08$ for $\alpha=2$. Such bounds on $\Upsilon$ are by now abundant in the literature, see for example \cite{Sakstein}, \cite{Jain}, \cite{Tapo3}, where
these were obtained from observational data. We emphasise however that the bounds discussed here are purely theoretical ones, and that
deriving these using observational data are more challenging in this context. Also, the bounds change slightly with varying stellar parameters. We also note
here that as shown by \cite{Maeder}, uniform stellar rotation leads to a further small decrease in $q_{max}$ and should make the bounds sharper. 
Next from the discussion of section \ref{ModelA}, we note that in the isotropic case, negative $\Upsilon$ leads to an increase in the SC limit and therefore an increase in the lifetime of 
the shell burning phase. For example, with $\alpha=2.0$ in the isotropic case, $\Upsilon=-0.25$ results in $32\%$ increase in $q_{max}$, which leads to an increased lifetime $t_{shell}^{*}$, which is $5.5$ times the lifetime $t_{shell}$ in the standard case. There is however no obvious way to put a theoretical lower bound on $\Upsilon$ from our analysis.} 

\tl{In an entirely similar manner, with $\Upsilon=0$, a positive $\tau$ value leads to a decrease in the SC limit, and thus the shell burning lifetime. Here we obtain an upper bound $\tau \lesssim 0.01$ for $\alpha=1$ and $\tau \lesssim 0.008$ for $\alpha=2$. These values are consistent with the ones we have chosen in Figure \ref{figure2}. A negative $\tau$ value, however, increases the shell burning lifetime due to an increase in the SC limit. For example, with $\Upsilon=0$ and with $\alpha=2.0$, a value of $\tau=-0.01$ leads to an increased shell burning lifetime, which is $\sim 2$ times the one in the standard case.}
	
\tl{While a decrease in the shell burning lifetime corresponds to a lower number of stars in the shell hydrogen burning phase, an increase in $t_{shell}$ indicates a higher number of stars in the shell burning phase, compared to the theoretical prediction in the standard case. In principle, such an increase in the population may have observational consequences.}

\section{Discussions}
\label{disc}

In this paper, we have presented results on the effects of modified gravity and anisotropy on the Sch\"{o}nberg-Chandrasekhar limit, 
which corresponds to the maximum fraction of a star's mass that can be in an isothermal core, supporting the overlying radiative envelope, 
\tl{and have indicated how one can obtain approximate bounds on the parameters of the theory from theoretical considerations of the SC limit.}
We adopted a model with an isothermal core and an $n=3$ polytropic envelope (model A). 
We have also carried out the analysis with a second model, namely one with 
an isothermal non-degenerate core, surrounded by a radiative envelope governed by Kramer's opacity law and a hydrogen 
burning shell at the core-envelope junction (model B). Importantly, radiation pressure is ignored in the latter. 
Without going into the details, we simply mention that the results are almost identical in the two cases. 
This can be gleaned from the isotropic case, for which we present the following comparison Table \ref{Table3}, which 
should make our argument clear. \tl{A more realistic analysis of the effects of anisotropy and modified gravity on the evolution of intermediate mass stars, 
using currently available stellar evolution codes, is left for a future study.}
\begin{table}
\begin{center}
\caption{Comparison of the SC limit between Model A and Model B in the isotropic case.}
\begin{tabular}{|c|c|c|c|c|}
\hline
$\Upsilon$ & \multicolumn{2}{c|}{Model A} &\multicolumn{2}{c|}{Model B} \\
\cline{2-5}
& $\alpha=1$ & $\alpha=2$ & $\alpha=1$ & $\alpha=2$\\
\hline
-0.25 & 0.425 & 0.104 & 0.438 & 0.132\\
\hline
0 & 0.359 & 0.079 & 0.371 & 0.101\\
\hline
0.25 & 0.306 & 0.063 & 0.317 & 0.082\\ 
\hline 
\end{tabular} 
\label{Table3}
\end{center}
\end{table}

\begin{figure}
\centering
\includegraphics[width=80mm]{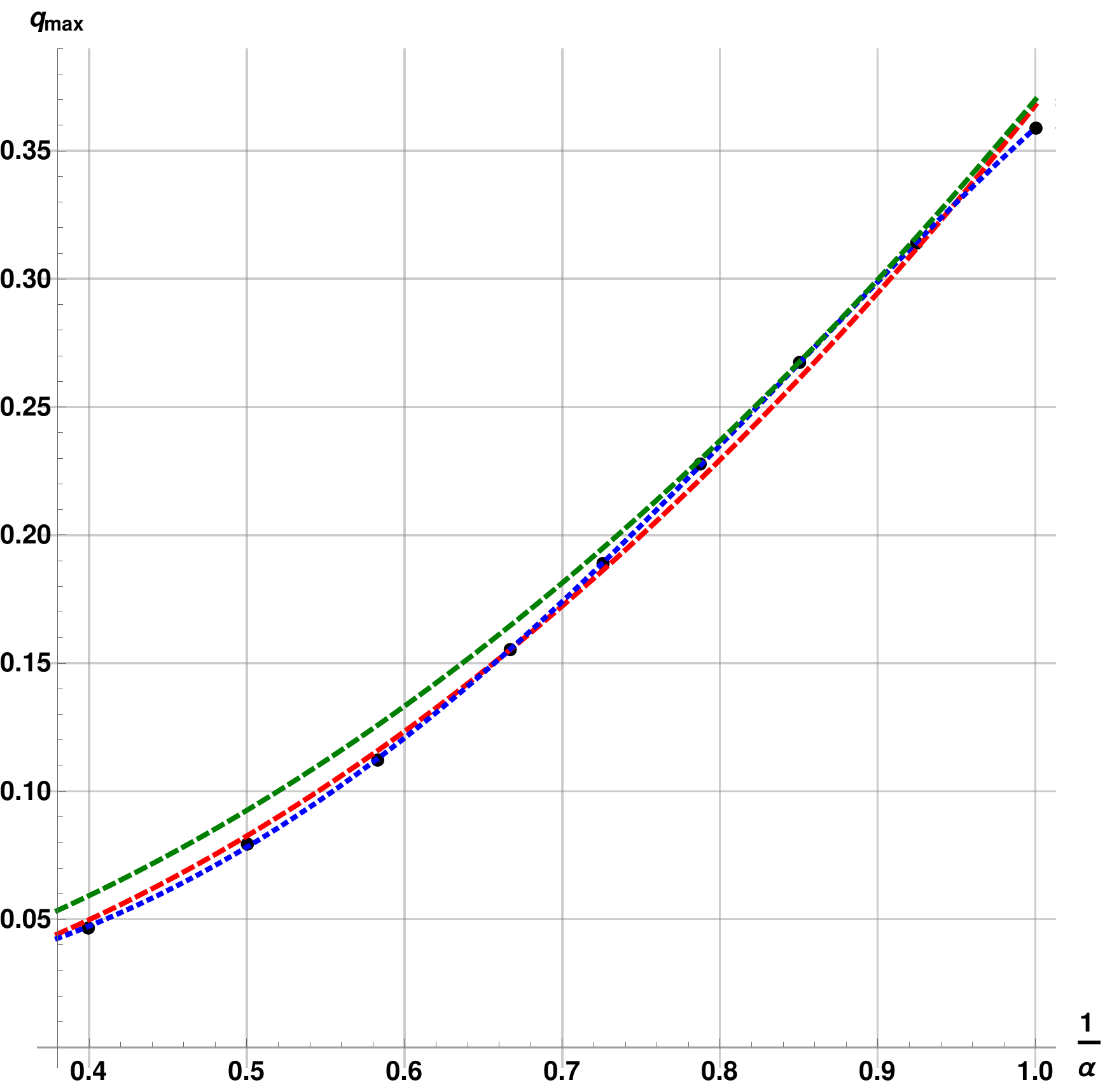}
\caption{\tred{Comparison between second order and fourth order best-fit formula, and the \tr{conventional} formula $q_{max}=0.37(1/\alpha)^2$
for the SC limit. The black dots correspond to numerically obtained data points. The green dashed curve represents the 
purely quadratic \tr{formula} for the 
conventional SC limit $q_{max}=0.37(1/\alpha)^2$. The blue dotted curve corresponds to the best-fit quartic formula 
obtained from the data points. The red dashed curve corresponds to the inhomogeneous quadratic formula obtained from the data points.}}
	\label{fig8}
\end{figure}

\tred{Our main result of this paper is deriving the quartic master formula for the SC limit as a function of $\Upsilon$, $\tau$ 
and $\alpha$, which is encoded in Figures \ref{figure2}(a) and (b). We also show a comparison of the numerically obtained formulae 
for the SC limit in the standard case, as compared to the \tr{conventional one}, see Figure \ref{fig8}. The green dashed curve is the usual formula for the SC limit in the standard case, mentioned in the introduction; \tr{it is a purely quadratic function in $1/\alpha$.} 
The blue dotted curve corresponds to the best-fit quartic formula derived from the numerically obtained data points for 
the standard case, which we find to be given by
\begin{equation}
q_{max}=0.128 \Big(\frac{1}{\alpha}\Big) - 0.500 \Big(\frac{1}{\alpha}\Big)^2 + 1.49 
\Big(\frac{1}{\alpha}\Big)^3 - 0.765 \Big(\frac{1}{\alpha}\Big)^4~.
\label{inhomoquartic}
\end{equation} 
The red dashed curve corresponds to the quadratic formula \tr{obtained from fitting the numerical data points for the standard case. It is an inhomogeneous quadratic function in $1/\alpha$, i.e., it contains a linear term in addition to a quadratic one. From Figure \ref{fig8}, it is observed that the quadratic formula is almost indistinguishable from the quartic one.}}
\tr{The main takeaway message from here is that a proper numerical analysis yields the result that the linear term is as significant as the quadratic term and thus should not be left out in the standard SC limit formula contrary to the textbook convention.} 

\tred{Ideally the formula one should obtain from the master formula under a certain limiting case, should be close to 
the best-fit formula obtained solely from the data points for that particular limiting case. Now it is 
observed that the isotropic limit of this master formula and the quartic fitting formula derived 
from the isotropic data points (section \ref{ModelA}), differs maximally by $1\%$. Furthermore, 
in the standard case, our master formula maximally differs by $1.3\%$ from the best-fit quartic fitting 
formula derived from the corresponding data points, Equation (\ref{inhomoquartic}). 
These percentages are significantly small when compared to the percentage changes in the SC limit due 
to either modified gravity or pressure anisotropy. Therefore, upon further minimizing this difference by considering 
even more number of data points in the analysis, our results and observations should not change significantly.
}

\section*{Acknowledgements}
\label{acknowledge}

\teal{We acknowledge the High Performance Computing (HPC) facility at IIT Kanpur, India, where the numerical computations were carried out. S.C. thanks 
Pritam Banerjee for useful discussions. \tr{S.C. also thanks Warrick Ball and Aneta Wojnar for helpful email correspondence on a draft of this paper.}} 

\section*{Data Availability}
\label{DataAvailability}

The data underlying this article will be shared on reasonable request to the corresponding author.

\appendix
\section{Relevant equations and formulae for the isotropic case}
\label{isotropic}
In this appendix, we list the formulae used in subsection \ref{ModelA}. We begin with 
the expressions for $U$ and $V$ :
\begin{align}
	&U=\frac{4\pi r^{3}\rho(r)}{M(r)}~, \nonumber \\
	&V=\frac{GM(r)\rho(r)}{rP(r)}+\frac{\Upsilon}{4}G\frac{r\rho(r)}{P(r)}\left(8\pi r\rho(r)+4\pi r^{2}\frac{d\rho(r)}{dr}\right)~.
\end{align}
Note that near the center, $U\to 3$ and $V\to 0$, and that near the stellar surface, $U\to 0$ and $V\to \infty$. 
The transformation to non-dimensional variables are made using 
\begin{equation}
r=xR,~P(r)=p(x)\frac{GM^{2}}{4\pi R^{4}},~M(r)=q(x)M,~T(r)=t(x)\frac{\mu m_{H}}{k}\frac{GM}{R},
\label{Trans1}
\end{equation}
\tred{where $p, q, t$ and $x$ are non-dimensional variables.}
In terms of these, the boundary conditions at the centre ($x=0$) are $q(0)=0, t(0)=t_{c}, p(0)=p_{c}$ and at the
stellar \tred{surface} ($x=1$) are $q(1)=1, t(1)=0, p(1)=0$. We \tred{then} make another set of transformations, \tred{adapted for the core} 
\begin{equation}
x=x^{*}x_{0}~,~p(x)=p^{*}(x^{*})p_{0}~,~q(x)=q^{*}(x^{*})q_{0}~,~t(x)=t^{*}(x^{*})t_{0}~
\label{Trans2}
\end{equation}
where the asterisked quantities are the new non-dimensional variables \tred{corresponding to the core, which we call the core variables}. The five constants with subscripts $0$ 
are chosen to satisfy the following conditions for an isothermal non-degenerate core :
\begin{equation}
\frac{q_{0}}{t_{0}x_{0}}=1~,~\frac{p_{0}x_{0}^{3}}{t_{0}q_{0}}=1~,~p_{0}=p_{c}~,~t_{0}=t_{c}~.
\end{equation}
The homology invariants in the core are:
\begin{equation}
U=\frac{p^{*}(x^{*})x^3}{q^{*}(x^{*})}~,~~V=\frac{q^{*}(x^{*})}{x^{*}}+\frac{\Upsilon}{4}x^{*}\frac{d^{2}q^{*}(x^{*})}{d{x^{*}}^2}~,
\label{UVcore}
\end{equation}
and those in the polytropic envelope are 
\begin{equation}
U=\frac{\xi^{3}\theta(\xi)^{n}}{\phi(\xi)}~~~,~~~V=-(n+1)\frac{\xi}{\theta(\xi)}\frac{d\theta(\xi)}{d\xi}~.
\label{UVenvelopeA}
\end{equation}
In the envelope, the non-dimensional variables $\theta$, $\phi$ and $\xi$ are used:
\begin{equation}
r=\eta \xi~~,~~\rho(r)=\rho_{c}\theta(\xi)^{n}~~,~~M(r)=4\pi \eta^{3}\rho_{c}\phi(\xi)
\label{Trans3}
\end{equation}
$\rho_{c}$ being the density at core-envelope junction and $\eta$, the length scale defined by: 
\begin{equation}
\eta^2=\frac{(n+1)K}{4\pi G}\rho_{c}^{\frac{1}{n}-1}
\end{equation}

\section{Relevant equations and formulae including \tred{anisotropy}}
\label{anisotropic}
The pressure balance equation in the isothermal core (in terms of the variables introduced in Appendix \ref{isotropic}) is
\begin{equation}
\frac{dp^{*}(x^{*})}{dx^{*}}=-\frac{\left(\frac{q^{*}(x^{*})p^{*}(x^{*})}{{x^{*}}^{2}}
	+\frac{\Upsilon}{2}{p^{*}(x^{*})}^2x^{*}-2\bar{\tau}{x_{0}}^{2}x^{*}p^{*}(x^{*})\right)}
{\left(1+\frac{\Upsilon}{4}p^{*}(x^{*}){x^{*}}^2\right)}~,
\label{HECAaniso}
\end{equation}
where $\beta(r)=\bar{\tau}(r/R)^2$, with $\bar{\tau}$ being the non-dimensional \tred{constant, quantifying the} 
anisotropy parameter in the stellar core, and $R$ is the radius of the star. Also,
\begin{equation}
x_0=\left(\frac{{T_c}^2k^2}{\mu^2{m_{H}}^2G4\pi R^2 P_c}\right)^{1/2}
\end{equation}
We also obtain the modified form of the homology \tred{invariant} variable
\begin{equation}
V=\frac{\left(\frac{q^{*}(x^{*})}{x^{*}}+\frac{\Upsilon}{2}{x^{*}}^2p^{*}(x^{*})-
	2\bar{\tau}{x_{0}}^{2}{x^{*}}^2\right)}{\left(1+\frac{\Upsilon}{4}p^{*}(x^{*}){x^{*}}^2\right)}~.
\label{Vcoreaniso}
\end{equation}
On the envelope side, the modified pressure balance equation is
\begin{equation}
\frac{d\theta(\xi)}{d\xi}=-\frac{\left(\frac{\phi(\xi)}{\xi^2}+\frac{\Upsilon}{2}
	\theta(\xi)^n\xi-2\tau\frac{\xi\theta(\xi)}{(n+1)}\right)}{\left(1+\frac{\Upsilon}{4}\theta(\xi)^{n-1}\xi^2n\right)}~,
\label{HECpolyAniso}
\end{equation}
where we have used $\beta(r)=\tau(r/\eta)^2$, with $\tau$ being the non-dimensional \tred{constant, quantifying the} 
anisotropy parameter in the stellar envelope.
Also, 
\begin{equation}
V=(n+1)\frac{\xi}{\theta(\xi)}\frac{\left(\frac{\phi(\xi)}{\xi^2}+\frac{\Upsilon}{2}
	\theta(\xi)^n\xi-2\tau\xi\frac{\theta(\xi)}{(n+1)}\right)}{\left(1+\frac{\Upsilon}{4}\theta(\xi)^{n-1}\xi^2n\right)}
\label{UVenvelopeAaniso}
\end{equation}	
The boundary conditions in the core and envelope respectively, remain unaltered compared to the isotropic case, 
with the only modification being in the definition of $\xi_0$, which is here defined as
\begin{equation}
\xi_{0}=\sqrt{\frac{V_{0}}{\Big[\frac{(n+1)}{U_{0}}+\Upsilon\left(\frac{(n+1)}{2}-\frac{nV_{0}}{4}\right)-2\tau\Big]}}
\label{startpointUVaniso}
\end{equation}
From the matching condition of the anisotropy parameter $\beta(r)$ at the
core-envelope junction, we obtain \teal{an} approximate relationship
$\bar{\tau}=\tau{\bar{\xi_{1}}}^{2}$, which estimates $\bar{\tau}$ for a given
$\tau$. $\bar{\xi_{1}}\simeq 6.9$ is the first zero of the polytrope (n = 3) in the
standard case.

\label{lastpage}
\end{document}